\documentclass[conference]{IEEEtran}
\usepackage{amsthm, amssymb, amsmath}
\usepackage{verbatim, float}
\usepackage{mathrsfs}
\usepackage{graphics}
\usepackage[usenames,dvipsnames]{pstricks}
\usepackage{epsfig}
\usepackage{pst-grad} 
\usepackage{pst-plot} 
\usepackage{cases}
\usepackage{algorithmic}
\usepackage{bm}
\usepackage[noadjust]{cite}
\usepackage[top=1.2cm, bottom=0.86in, left=0.6in, right=0.6in]{geometry}
\usepackage{url}

\usepackage{hhline}

\usepackage{etoolbox}
\makeatletter
\patchcmd{\@makecaption}
  {\scshape}
  {}
  {}
  {}
\makeatother

\usepackage[singlelinecheck=false]{caption}
\usepackage{diagbox}

\theoremstyle{plain}
\newtheorem{theorem}{Theorem}
\newtheorem{lemma}{Lemma}

\newtheorem{corollary}{Corollary}

\theoremstyle{definition}
\newtheorem{definition}{Definition}
\newtheorem{example}{Example}

\newcommand{\C}{{\mathcal C}}
\newcommand{\D}{{\mathcal D}}

\newcommand{\F}{{\mathcal F}}

\newcommand{\T}{{\mathcal T}}

\renewcommand{\S}{{\mathcal S}}

\newcommand{\Sds}{{\mathcal S}_{\delta}^*}

\newcommand{\ba}{{\boldsymbol a}}

\newcommand{\bc}{{\boldsymbol c}}

\newcommand{\bcj}{{\boldsymbol c}_j}

\newcommand{\bco}{{\boldsymbol c}_1}
\newcommand{\bcn}{{\boldsymbol c}_n}
\newcommand{\bct}{{\boldsymbol c}_2}

\newcommand{\bbF}{{\mathbb F}}

\newcommand{\bbZ}{{\mathbb Z}}

\newcommand{\ff}{\mathbb{F}}

\newcommand{\fq}{\mathbb{F}_q}

\newcommand{\fql}{\mathbb{F}_{q^{\ell}}}
\newcommand{\fqls}{\mathbb{F}_{q^{\ell}}^*}

\newcommand{\fqm}{\mathbb{F}_{q^m}}

\newcommand{\bal}{\bm{\alpha}}
\newcommand{\bals}{\bm{\alpha}^*}

\newcommand{\bla}{\bm{\lambda}}

\newcommand{\spn}{\mathsf{span}_{\fq}}

\newcommand{\rank}{\mathsf{rank}_{\fq}}

\newcommand{\mhs}{\mathsf{MHS}}
\newcommand{\base}{\mathsf{base}}

\newcommand{\vbc}{\vec{\boldsymbol{c}}}
\newcommand{\vc}{\vec{\boldsymbol{c}}}

\newcommand{\vg}{\vec{\boldsymbol{g}}}

\newcommand{\vlam}{\vec{\boldsymbol{\lambda}}}

\newcommand{\define}{\stackrel{\mbox{\tiny $\triangle$}}{=}}

\newcommand{\rsk}{\text{RS}(A,k)}
\newcommand{\grskl}{\text{GRS}(A,k,\vec{\boldsymbol{\lambda}})}
\newcommand{\grsnkl}{\text{GRS}(A,n-k,\vec{\boldsymbol{\lambda}})}

\newcommand{\faj}{f(\boldsymbol{\alpha}_j)}

\newcommand{\gaj}{g(\boldsymbol{\alpha}_j)}

\newcommand{\bw}{{\sf{bw}}}
\newcommand{\stabS}{{\sf{Stab}}_{\fqls}(S^*)}
\newcommand{\nqld}{N_{q,\ell,\delta}}

\begin{document}

\title{Designing Compact Repair Groups\\for Reed-Solomon Codes
\thanks{
T. X. Dinh, Serdar Boztas, and S. H. Dau are with the School of Computing Technologies, RMIT University, Australia. Emails: \{S3880660,serdar.boztas, sonhoang.dau\}@rmit.edu.au.  
T. X. Dinh is also with the Department of Mathematics, Faculty of Natural Science and Technology, Tay Nguyen University, Vietnam.
Emanuele Viterbo is with the Department of Electrical Engineering, Monash University. Email: emanuele.viterbo@monash.edu.}}




%
%
   
%
 \author{%
   \IEEEauthorblockN{Thi Xinh Dinh\IEEEauthorrefmark{1}\IEEEauthorrefmark{2},
                     Serdar Boztas\IEEEauthorrefmark{1}\IEEEauthorrefmark{3},
                     Son Hoang Dau\IEEEauthorrefmark{1},
                     and Emanuele Viterbo\IEEEauthorrefmark{4}}
   \IEEEauthorblockA{\IEEEauthorrefmark{1}%
                     RMIT University,
                        Melbourne VIC 3000, Australia,
                     \{s3880660@student.rmit, serdar.boztas@rmit, sonhoang.dau@rmit\}.edu.au}
   \IEEEauthorblockA{\IEEEauthorrefmark{2}%
                     Tay Nguyen University, Vietnam,  dinhthixinh@ttn.edu.vn}
                     
   \IEEEauthorblockA{\IEEEauthorrefmark{3}%
                     University Research Foundation, Greenbelt, MD 20770, USA, boztas@urf.com}
   \IEEEauthorblockA{\IEEEauthorrefmark{4}%
                     Monash University, Clayton VIC 3800, Australia,                emanuele.viterbo@monash.edu}
}

\maketitle

\maketitle

\begin{abstract}
Motivated by the application of Reed-Solomon codes to recently emerging decentralized storage systems such as Storj and Filebase/Sia, we study the problem of designing \textit{compact} 
repair groups 
for recovering multiple failures in a decentralized manner.
Here, compactness means that the corresponding trace repair schemes of these groups of helpers can be generated from a single or a few \textit{seed} repair schemes, thus saving the time and space required for finding and storing them.
\textit{The goal} is to design compact repair groups that can tolerate as many failures as possible.
It turns out that the maximum number of failures a collection of repair groups can tolerate equals the size of a minimum hitting set of a collection of subsets of the finite field $\fql$ minus one. When the repair groups for each symbol are generated from a \textit{single} subspace, we establish a pair of asymptotically tight \emph{lower bound and upper bound} on the size of such a minimum hitting set. 
Using Burnside's Lemma and the M\"{o}bius inversion formula, we determine 
a number of subspaces that together attain the upper bound on the minimum hitting set size when the repair groups are generated from \textit{multiple} subspaces.
\end{abstract}


\section{Introduction}
\label{sec:intro}

Unlike traditional distributed storage systems such as Google File System II~\cite{colossus_2010} or Facebook's f4~\cite{FBf42014}, in which most failures among the servers storing a file or a data object are \textit{single failures}~\cite{Dimakis_etal2010}, in recently emerging \textit{decentralized} storage systems such as the blockchain-based Storj~\cite{Storj} and Filebase/Sia~\cite{Filebase}, \textit{multiple failures} are the norm. This difference between the distributed and the decentralized systems stems from the fact that storage nodes in a decentralized system are not under the control of any centralized party and can join and leave the system or go online and offline more freely. By contrast, in traditional distributed storage systems, all storage nodes are managed by a centralized service provider and rarely fail or leave the system.

As a consequence, Reed-Solomon codes employed in decentralized systems have significantly different parameter ranges. For example, while most major distributed storage systems use short codes with three or four parities, e.g., RS(9,6) or RS(14,10), decentralized storage systems like Storj and Filebase/Sia rely on longer codes with larger redundancies, e.g., RS(30,10), RS(40,20), or RS(80,40)~\cite[Table~I]{DinhNguyenMohanSerdarLuongDau_ISIT2022}. 
Larger redundancies allow these codes to tolerate a greater number of node failures as often occurring in a decentralized environment.

\begin{figure}[htb]\label{fig1}
\centering
\includegraphics[scale=0.60]{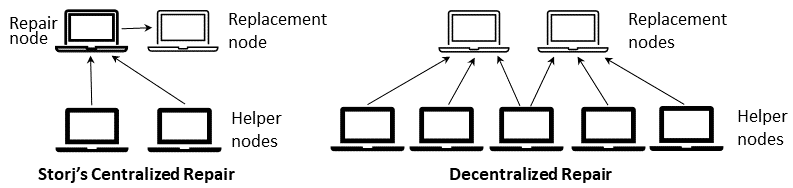}
\caption{An illustration of Storj's centralized approach and the decentralized approach when repairing $e=2$ failed nodes in a code with dimension $k=2$.}
\label{tab:DSS}
\vspace{-15pt}
\end{figure}

When nodes storing codeword symbols corresponding to a particular data object fail (or leave the system), other nodes may join as replacement nodes for that object. These can be new nodes that just joined the system or existing nodes that are ready to provide extra storage. There must be a mechanism for these nodes to recover the lost symbols. 
Current decentralized systems like Storj~\cite{Storj} are using the ``lazy repair'' mechanism~\cite{TotalRecall_NSDI_2004, Silberstein2014, LubyPadovaniRichardsonMinderAggarwal_ACMTS_2019}, which waits for a number of failures to happen before applying the \textit{naive centralized repair} approach to repair all failures at once.
This approach assigns one random node as the \textit{repair node}, which collects codeword symbols from $k$ arbitrary available nodes ($k$ is the code dimension), recovers all lost codeword symbols at once and sends them to the corresponding replacement nodes (see Fig.~\ref{tab:DSS}). 
We observe that the \textit{repair bandwidth} of the naive centralized repair approach, i.e., the total amount of bits communicated among the storage nodes during the repair process, can potentially be reduced by using the trace repair method for multiple failures developed in the literature~\cite{GuruswamiWootters2016, GuruswamiWootters2017, DauDuursmaKiahMilenkovic2018,BartanWootters2017, MardiaBartanWootters2018, LiWangJafarkhani_TIT_2019, TamoYeBarg2019}.
Also, as the (heuristic) search problem for a low-bandwidth repair scheme is highly intensive for multiple failures, algebraic constructions offer a more feasible approach, which, however, work for a limited range of parameters only.

We study in this work a second approach, namely, the \textit{decentralized repair} approach, in which different replacement nodes perform the repair process independently not necessarily simultaneously.
In its naive version, each replacement node performs the naive repair scheme, downloading data from $k$ helper nodes and recovering the lost symbol.
We are interested in a more bandwidth-efficient version in which the trace repair method~\cite{GuruswamiWootters2016, GuruswamiWootters2017} is applied: each replacement node downloads data from more than $k$ nodes, which, however, results in lower repair bandwidth compared to the naive scheme.
While this approach may require a higher total repair bandwidth compared to Storj's naive \textit{centralized} repair approach\footnote{To repair $e$ failed nodes for an RS$(n,k)$ code, Storj's repair scheme requires a total bandwidth of $k\ell + (e-1)\ell$ symbols, while our scheme requires $ek\ell(1-s)$ symbols, where $s$ represents the saving/reduction in bandwidth achieved by the trace repair approach compared to the naive one, e.g., $s = 0.3$.}, 
the recovery task is split over multiple nodes, which eliminates the communication and computation bottle neck at the single repair node. This may potentially help to reduce the total recovery time of the system and lower the chance that the recovery process is disrupted due to the failure of the assigned repair node itself. 
Moreover, repairing nodes actively instead of lazily reduces the risk of data loss due to correlated failures (software bugs or virus attacks).

However, using trace repair to improve the repair bandwidth incurs additional overheads: the computational overhead when finding a trace repair scheme (see~\cite[Section IV.~B]{DinhNguyenMohanSerdarLuongDau_ISIT2022}) and the storage overhead to store the parameters of a trace repair scheme (see~\cite[Section V.~B]{DinhNguyenMohanSerdarLuongDau_ISIT2022}). 
One extreme solution that achieves minimum computation/storage overhead is to use a \textit{single} repair group of fixed $d$ helper nodes for each codeword symbol. Unfortunately, this solution cannot tolerate even one failure: as soon as one node in that repair group fails, the corresponding repair scheme no longer works. At the other extreme, 
one can tolerate up to $n-1-d$ failures by selecting \textit{all} $\binom{n-1}{d}$ different repair groups and find/store the corresponding repair schemes. Clearly, this solution requires an exponential overhead in computation and storage. \textit{Our goal} is to design \textit{compact} repair groups for every codeword symbol, which incur low computational and storage overhead while tolerating as many failures as possible. Within the scope of this work, we only consider full-length codes over $\fql$, with length $n = q^\ell$.

Our key insight for 
constructing compact repair groups is as follows: 
instead of using many \textit{unrelated} repair groups (resulting in high computation/storage overheads), 
we carefully select \textit{related} repair groups whose corresponding repair schemes can be interpolated from a single or a few ``seed'' repair schemes to reduce the overhead of constructing and storing them. 
More specifically, to repair a codeword symbol $c_j$, we pick a $d$-subset $S \owns 0$, referred to as a \textit{seed}, and use all of its unique multiplicative cosets $\C(S^*) \define \big\{b S^* \colon b\in \fqls\big\}$, where $S^* = S \setminus \{0\}$ and $\fqls = \fql \setminus \{0\}$, as different repair groups (after an appropriate shifting). It turns out that the repair schemes for all these groups can be easily generated from a single repair scheme corresponding to $S^*$. Moreover, the maximum number of failures tolerable by these compact repair groups is equal to the size of a minimum hitting set of $\C(S^*)$ minus one. 
Finally, to further improve this quantity, we can start with multiple seeds.
To achieve tractable results, we study an $\fq$-subspace $S$ of $\fql$ and consider full-length codes.

Our contributions are summarized below. 
\vspace{-3pt}
\begin{itemize}
    \item First, we introduce a new problem of designing compact repair groups for Reed-Solomon codes in the decentralized setting and connect it to the minimum hitting set problem.
    \item Second, when the repair groups are generated from a single seed $S$, we establish \textit{asymptotically matching lower bound and upper bound} on the size of a minimum hitting set of $\C(S^*)$ (as $q\to \infty$). Note that in general, the best approximation ratio for the size of a minimum hitting set of an arbitrary collection of sets is in the order of $\log(n)=\log(q^\ell)=\ell\log(q)$~\cite{LundYannakakis1994}, \cite{Feige1998}.
    \item Lastly, we show that the upper bound on the number of tolerable failures can be attained by using $s$ seeds where $s$ is the number of orbits of $\S^*_\delta$, the set of all $S^*$, where $S$'s are $\delta$-dimensional $\fq$-subspaces of $\fql$, $\delta = \log_qd$, under the action of the multiplicative group $\fqls$. We derive an explicit formula for $s$ via Burnside's Lemma and the M\"{o}bius inversion formula.
\end{itemize}


\section{Preliminaries}
\label{sec:pre}

\subsection{Definitions and Notations}
\label{subsec:def-notation}
\vspace{-2pt}
Let $q$ be a prime power, $\fq$ be the finite field of $q$ elements and $\fql$ be the extension field of degree $\ell$ of $\fq$.
We use $[n]$ to denote the set $\{1, 2, \ldots,n\}$, and $a\mid b$ to denote that $a$ divides $b$, for two integers $a$ and $b$. For a set $S$, let $S^* \define S\setminus \{0\}$.

Let $C$ be a \emph{linear} $[n,k]$ \emph{code}  over $\fql$. Then $C$ is an $k$-dimensional $\fql$-subspace of $\fql^n$. A \emph{codeword} of $C$ is an element $\vbc=(\bco,\bct,\ldots,\bcn)\in C$ and its codeword symbols are the components $\bcj$, $j \in [n]$. 
The \emph{dual} code of a code $C$ is the orthogonal complement $C^\perp$ of $C$, $C^\perp = \{\vg \in \fql: \langle \vc,\vg\rangle = 0, \forall \vc \in C\}$, where $\langle \vc, \vg \rangle$ is the scalar product of $\vc$ and $\vg$. The code $C^\perp$ is an $\fql$-subspace of $\fql^n$ with dimension $n - k$.
The elements of $C^\perp$ are called \textit{dual codewords}. 
The number $n-k$ is called the \emph{redundancy} of the code $C$.

\begin{definition} 
\label{def:RS}
Let $A=\{\bal_j\}_{j=1}^n$ be a subset size $n$ in $\fql$. A \textit{Reed-Solomon code} $\rsk \subseteq \fql^n$ of dimension $k$ with \textit{evaluation points} $A$ is defined as \vspace{-5pt} 
\[
\rsk = \big\{\big(f(\bal_1),\ldots,f(\bal_n)\big) \colon f \in \fql[x],\ \deg(f) < k \big\}, \vspace{-5pt}
\]
where $\fql[x]$ is the ring of polynomials over~$\fql$.
We also use the notation RS$(n,k)$, ignoring the evaluation points. 
\end{definition}
\vspace{-5pt}
A \emph{generalized} Reed-Solomon code, $\grskl$, where $\vec{\bla} = (\bla_1,\ldots,\bla_n)\in \fql^n$, is the set of codewords $\big( \bla_1g(\bal_1),\ldots,\bla_n g(\bal_n) \big)$, where $\bla_j \neq 0$ for all $j \in [n]$, $g \in \fql[x],\ \deg(g) < n-k$.
The dual code of a Reed-Solomon code $\rsk$ is a generalized Reed-Solomon code $\grsnkl$, 
for some multiplier vector $\vec{\bla}$~(\cite[Chap.~10]{MW_S}). 
We sometimes use the notation GRS$(n,k)$, ignoring $A$ and $\vlam$.

Let $f(x)$ be a polynomial corresponding to a codeword of the Reed-Solomon code $C=\rsk$, and $g(x)$ be a polynomial of degree at most $n-k-1$, which corresponds to a codeword of the dual code $C^\perp$. Then
$
\sum_{j=1}^n \gaj\big(\bla_j\faj\big) = 0, 
$
and we call the polynomial $g(x)$ a \textit{check} polynomial for $C$. 

For a subset $S \subseteq \fql$ and $a \in \fql$, let 
$aS \triangleq \{as\colon s\in S\}$,
$a + S \triangleq \{a+s\colon s\in S\}$,
$\C(a, S) \triangleq \{a+bS: b\in \fqls\}$,
and
$\C(S) \triangleq \{bS: b \in \fqls\}$.
Let $\S$ be a family of sets $\{S_i\}_{i \in I}$. A \textit{hitting set} of $\S$ is a set which has nonempty intersection with $S_i$, for all $i\in I$. A hitting set of $\S$ is called a \textit{minimum hitting set} (MHS), if it has the smallest size over all the hitting sets of $\S$. For simplicity, we denote by $\mhs(\S)$ a MHS of $\S$. 

\vspace{-3pt}
\subsection{Trace Repair Method}
\label{subsec:tracerepair}
\vspace{-3pt}

Let RS$(n,k)$ be a Reed-Solomon code over $\fql$ with evaluation points $A$,
$\vc$ ~a codeword corresponding to polynomial $f(x)$, $f \in \fql[x],\ \deg(f) < k$, and $\bc^* = f(\bals)$, where $\bals \in A$, a codeword symbol/node of $\vc$.
A (linear) \emph{trace repair scheme} for $\bc^*$ corresponds to a set of $\ell$ check polynomials $\left\{g_i(x)\right\}_{i\in [\ell]}$, $g_i \in \fql[x],\ \deg(g_i) < n-k$, that satisfies the \textit{Full-Rank Condition}:
$\rank\{g_i(\bals)\}_{i\in [\ell]}=\ell$.
If we require that the set of helper nodes corresponds to $S\setminus \{\bals\}$ where $\bals \in S\subseteq A$, then 
it is necessary that $g_i(\bal) = 0$ for every $i \in [\ell]$ and $\bal \notin S$. The repair bandwidth of such a repair scheme (in $\fq$-symbols) is ${\sf{bw}} = \sum_{\bal\in S \setminus \{\bals\}} \rank(\{g_i(\bal)\}_{i \in [\ell]})$. 

To repair all $n$ components of~$\vc$, we need $n$ such repair schemes (possibly with repetition). 
See, e.g. \cite{DauDinhKiahTranMilenkovic2021}, for a detailed explanation of why the above scheme works with an example. 



\section{Designing Compact Repair Groups for Reed-Solomon Codes}
\label{sec:constructions}
\vspace{-5pt}

\subsection{The Problem Description}
\label{sub:problem}
\vspace{-5pt}

\textbf{The context}. We assume that a full-length Reed-Solomon code $RS(n,k)$ is employed in a decentralized storage system where multiple failures occur. A number of storage nodes, i.e., codeword symbols, are lost and need to be recovered. To repair each failed node, the corresponding replacement node will contact a set of $d \leq n-1$ helper nodes among the surviving nodes (independently of other replacement nodes). A low-bandwidth repair scheme corresponding to each of such group of $d$ helper nodes and the lost node must be determined and stored before it can be used in the repair process.

\textbf{The problem}. The trace repair method~\cite{GuruswamiWootters2016,GuruswamiWootters2017} can be employed to construct repair schemes with low bandwidths. 
As it may take a significant time to find a low-bandwidth repair scheme with respect to an arbitrary repair group via heuristic algorithms (see~\cite{DinhNguyenMohanSerdarLuongDau_ISIT2022}), the system should find all repair schemes offline and pre-store their parameters. The more repair groups are used, the larger the number of failures that these groups can tolerate (i.e. at least one group is intact).
However, having more repair groups requires higher computation/storage cost. We aim to address this dilemma in this work.
\vspace{-6pt}

\subsection{The General Strategy}
\label{subsec:general}
\vspace{-5pt}

Our key idea is to use multiple \textit{related} repair groups, i.e. although multiple repair schemes will be used, they are all generated from one or a few seed repair schemes only. We refer to these as \textit{compact} repair groups. For such repair groups, we only need to store one or a few sets of evaluation points, referred to as \textit{seeds}, and their corresponding repair schemes, which can then be used to generate all others.

More specifically, for the case of a single seed, assume that we already have a repair scheme for $f(0)$ with helper nodes correspond to $S^*\define S\setminus \{0\}$, where $0 \in S\subseteq \fqls$. Then for every $\bals \in \fql$, we can create a repair scheme for $f(\bals)$ with the same bandwidth and with helper nodes correspond to $\bals + bS^*$ for every $b \in \fqls$. As shown in Lemma~\ref{lem:translating-dilating} (see also \cite[Corollary 4]{DinhNguyenMohanSerdarLuongDau_ISIT2022}), this can be done by appropriately dilating and translating the (seed) repair scheme for $f(0)$ to yield a repair scheme for $f(\bals)$. 
For multiple seeds, we simply start from multiple sets $\{S_t\}_{t \in [s]}$. Note that all these statements hold under the assumption that the considered RS code is full length, so that $\bals + bS^*$ corresponds to a subset of evaluation points, i.e. a valid set of helper nodes, for every $\bals$, $b$, and $S$.\vspace{-5pt}

\begin{lemma}
\label{lem:translating-dilating}
Assume that $0 \in S\subseteq \fql$. 
Suppose that $\{g_i(x)\}_{i=1}^\ell \subset \fql[x]$ forms a repair scheme with bandwidth $\bw$ and helper nodes correspond to $S^*$ for $f(0)$ of a RS code with evaluation points $S$. Then $\{h_i(x)\define g_i\big((x-\bals)/b\big)\}_{i=1}^\ell$ forms a repair scheme with bandwidth $\bw$ and helper nodes correspond to $\bals + bS^*$ for $f(\bals)$ of a RS code of the same dimension with evaluation points $\bals + bS$, for every $b\in \fqls$.
\end{lemma} 
\begin{proof}
Since $h_i(\bals) = g_i(0)$ for every $i \in [\ell]$, we have $\rank\big(\{h_i(\bals)\}_{i=1}^\ell\big) = \ell$, which means that the Full-Rank Condition is satisfied.
Furthermore, for $\ba \in \bals + bS^* = (\bals+bS)\setminus \{\bals\}$, we have $\ba = \bals + b\bal$ for some $\bal \in S^*$. Hence,
$h(\ba) = h(\bals + b\bal) = g(\bal)$ and as a consequence,
\[
\begin{split}
    \sum_{\ba \in \bals+bS^*}\hspace{-12pt}
\rank\big(\{h_i(\ba)\}_{i=1}^\ell\big)
= \sum_{\bal \in S^*} \hspace{-2pt}
\rank\big(\{g_i(\bal)\}_{i=1}^\ell\big)
= \bw. 
\end{split}
\] 
Thus, $\{h_i(x)\}_{i=1}^\ell$ forms a repair scheme for $f(\bals)$ with the same bandwidth as the repair scheme $\{g_i(\bal)\}_{i=1}^\ell$ for $f(0)$.
\end{proof}
\vspace{-7pt}

Let $\C(S^*) \define \{bS^*: b \in \fqls\}$, the set of distinct multiplicative cosets of $S^*$, and $\C(\bals, S^*) \define \{\bals+bS^*: b \in \fqls\}$. 
From the above discussion, the number of repair schemes for $f(\bals)$ generated from a single seed $S$ is $|\C(\bals, S^*)|=|\C(S^*)|$.
When node failures occur, all the repair schemes involving at least one failed node as a helper node will become unusable. We say that these repair schemes are \textit{hit} by those node failures. Hence, the maximum number of failures tolerable by these repair groups is one less than the size of a minimum hitting set of $\C(\bals, S^*)$ as well as of $\C(S^*)$. We summarize this discussion in Definition~\ref{def:max-failures} and Lemma~\ref{lem:max-failures}, generalized to capture the multiple-seed scenario.\vspace{-5pt}

\begin{definition}
\label{def:max-failures}
Let $\C\big(\bals, \{S^*_t\}_{t \in [s]}\big)$ denote the collection $\cup_{t \in [s]} \{\bals+bS_t^*: b \in \fqls\}$ for $s$ seeds $\{S_t\}_{t \in [s]}$. Then the maximum number of failures tolerable by this collection, denoted by $\F\big(\C\big(\bals, \{S^*_t\}_{t \in [s]}\big)\big)$, is defined to be the largest number $e$ so that for \textit{every} $E \subset \fql \setminus \{\bals\}$, $|E|=e$, there exists at least one set in the collection that doesn't intersect $E$. 
\end{definition}
\vspace{-7pt}

\begin{lemma}
\label{lem:max-failures}
Let $\{S_t\}_{t \in [s]}$ be a collection of $s$ subsets of $\fql$ containing $0$.
Then $\F\big(\C\big(\bals, \{S^*_t\}_{t \in [s]}\big)\big) = \left|\mhs\big(\C\big(\bals, \{S^*_t\}_{t \in [s]}\big)\big)\right|-1 = \left|\mhs\big(\C\big(\{S^*_t\}_{t \in [s]}\big)\big)\right|-1$, where $\C\big(\{S^*_t\}_{t \in [s]}\big) \define \cup_{t \in [s]} \{bS_t^*\colon b \in \fqls\}$.
\end{lemma}
\begin{proof}
Clearly, $\left|\mhs\big(\C\big(\bals, \{S^*_t\}_{t \in [s]}\big)\big)\right|$ is the minimum number of failures that hit all the repair groups in $\C\big(\bals, \{S^*_t\}_{t \in [s]}\big)$.
It is also straightforward that  $\left|\mhs\big(\C\big(\bals, \{S^*_t\}_{t \in [s]}\big)\big)\right| =  \left|\mhs\big(\C\big(\{S^*_t\}_{t \in [s]}\big)\big)\right|$.
Therefore, the repair groups corresponding to $\C(\bals, S^*)$ can tolerate at most $\left|\mhs\big(\C\big(\bals, \{S^*_t\}_{t \in [s]}\big)\big)\right|-1$ failures. 
\end{proof}
\vspace{-5pt}

Thanks to Lemma~\ref{lem:max-failures}, we have reduced the problem of determining the number of failures tolerable by a compact collection of repair groups generated by the seeds $\{S_t\}_{t \in [s]}$ to the problem of finding a minimum hitting set of $\C\big(\{S^*_t\}_{t \in [s]}\big)$.

\vspace{-3pt}
\subsection{Compact Repair Groups from a Single Seed}
\vspace{-3pt}

Finding an MHS of an arbitrary collection of sets is in general an NP-hard problem \cite{KorteVygen2012}. We focus on a special case when the seeds are $\fq$-subspaces of $\fql$. 
Note that low-bandwidth repair schemes for Reed-Solomon codes with evaluation points forming a subspace have been proposed in the literature~\cite{LiWangJafarkhani_TIT_2019, BermanBuzagloDorShanyTamo_ISIT_2021}. In this section we study $\left|\mhs\big(\C(S^*)\big)\right|$, where $S$ is a single seed and also a $\delta$-dimensional $\fq$-subspace of $\fql$. We establish a pair of lower bound and upper bound for $\left|\mhs\big(\C(S^*)\big)\right|$, which are asymptotically tight (Theorem \ref{theo:MHS-bounds}) as $q \to \infty$. Note that this is impossible to achieve in the general case. 
As a consequence, exact values for the MHS can be determined for some special cases (Corollary~\ref{cor:mhs-subfields}, Corollary~\ref{cor:ell-minus-one-dim}).

For the upper bound, for a $\delta$-dimensional $\fq$-subspace $S$ of $\fql$, we first bound $\left|\mhs\big(\C(S^*)\big)\right|$ by $\left|\mhs(\Sds)\right|$ where $\Sds$ is the collection of all $\delta$-dimensional $\fq$-subspaces of $\fql$ excluding $0$, and then connect $\mhs(\Sds)$ to a special case of Tur\'an designs~\cite{EtzionVardy2011}.
\vspace{-5pt}

\begin{lemma}
\label{lem:sds} If $S$ is a $\delta$-dimensional $\fq$-subspace $S$ of $\fql$ and $\Sds \define \{T^*\colon T \text{ is a } \delta\text{-dimensional } \fq\text{-subspace of } \fql\}$, then
$\left|\mhs\big(\C(S^*)\big)\right| \leq \left|\mhs\big(\Sds)\big)\right|$.
\end{lemma}
\begin{proof}
As a multiplicative coset of an $\fq$-subspace of $\fql$ is another $\fq$-subspace of the same dimension, if $\dim_{\fq}(S) = \delta$ then $\C(S^*) \subseteq \Sds$. The lemma follows.
\end{proof}

It turns out that a MHS of $\Sds$ corresponds to a special Tur\'an design~\cite{EtzionVardy2011}.
We recall below the definition of Tur\'an designs and discuss the relation with the MHS of $\Sds$.
\vspace{-3pt}
\begin{definition}
\label{def:turan-design}
Let $\S_\delta$ and $\S_r$ be the sets of all $\fq$-subspaces of $\fql$ of dimension $\delta$ and $r$, respectively. A \textit{$(q,\ell, \delta, r)$-Tur\'an design} is a subset $\D$ of $\S_r$ so that each subspace in $\S_\delta$ contains at least one subspace in $\D$.
The minimum size of a Tur\'an design is called the \textit{$(q,\ell, \delta, r)$-Tur\'an number} and denoted by $\T_q(\ell, \delta, r)$.
\end{definition}
\vspace{-5pt}
In Lemma~\ref{lem:turan-number-vs-mhs}, we show that $|\mhs(\Sds)|$ is the same as the  Tur\'an number (with $r=1$) defined above.

\begin{lemma}
\label{lem:turan-number-vs-mhs}
For every prime power $q$ and every $1 \leq \delta \leq \ell$, we have
$|\mhs(\Sds)| = \mathcal{T}_q(\ell, \delta, 1)$, 
where $\Sds$ is the set of all $\fq$-subspaces of dimension $\delta$ of $\fql$ (excluding the zero element) and $\mathcal{T}_q(\ell, \delta, 1)$ denotes the Tur\'an number in Definition~\ref{def:turan-design}.
As a consequence, $|\mhs(\Sds)| = \big(q^{\ell - \delta + 1}-1\big)/(q-1)$.
\end{lemma}
\begin{proof}
First, we show that for every hitting set $H$ of $\Sds$, there is a $(q, \ell, \delta, 1)$-Tur\'an design $\D_H$ 
of size $|\D_H| \leq |H|$.
Indeed, let $\D_H$ be the set of all (distinct) $1$-dimensional $\fq$-subspaces generated by elements in $H$. 
Since each $S^* \in \Sds$ contains at least one element $h \in H$, $S$ contains the corresponding $1$-dimensional subspace $\spn(\{h\})$.
Therefore, $\D_H$ is an $(q, \ell, \delta, 1)$-Tur\'an design with size $|\D_H| \leq |H|$.

Second, we show that if $\D$ is a $(q, \ell, \delta, 1)$-Tur\'an design then we can construct a hitting set $H$ of $\S^*_\delta$ of size $|H| = |\D|$.
Indeed, let $H$ be the set of generators of all $1$-dimensional subspaces in $\D$. Then $|H|= |\D|$ and $H$ hits all the set $S^* \in \Sds$.  

Thus,
$|\mhs(\Sds)| = \T_q(\ell, \delta, 1) =
\big(q^{\ell - \delta + 1}-1\big)/(q-1)$ (see~\cite[Theorem $4.10$]{EtzionVardy2011}).
\end{proof}

An upper bound on $\left|\mhs\big(\C(S^*)\big)\right|$ can be directly derived from Lemma~\ref{lem:sds} and Lemma~\ref{lem:turan-number-vs-mhs}.
\begin{corollary}
\label{cr:UB}
Let $S$ be a $\delta$-dimensional $\fq$-subspace of $\fql$. 
Then $\left|\mhs\big(\C(S^*)\big)\right| \leq \big(q^{\ell - \delta + 1}-1\big)/(q-1)$.
\end{corollary}

We now discuss a lower bound on $\left|\mhs\big(\C(S^*)\big)\right|$.
\begin{lemma}
\label{lem:LB}
Let $S$ be a $\delta$-dimensional $\fq$-subspace of $\fql$. 
Then $\left|\mhs\big(\C(S^*)\big)\right| \geq \left\lceil\frac{q^\ell - 1}{q^\delta -1}\right\rceil$.
\end{lemma}
\begin{proof}
We first note that every two elements $x$ and $y$ of $\fqls$ belong to the same number of cosets in $\C(S^*)$ because if $x \in bS^*$, then  $y = \frac{y}{x}x \in \big(\frac{y}{x}b\big)S^*$ and vice versa.
By double counting the set $\left\{(x,bS^*)\colon x \in bS^*, b \in \fqls\right\}$, each element in $\fqls$ appears in exactly $r \define \frac{|\C(S^*)||S^*|}{q^\ell -1}$ cosets, noting that $|bS^*| = |S^*|$ for every $b \in \fqls$.
Now, suppose that $H$ is a hitting set of $\C(S^*)$. As proved above, the elements in $H$ together hit at most $r|H|$ cosets. This implies that $r|H| \geq |\C(S^*)|$, which in turn shows that $|H| \geq \left\lceil\frac{q^\ell - 1}{q^\delta -1}\right\rceil$, establishing the lemma.
\end{proof}


The derived lower bound and upper bound on $\left|\mhs\big(\C(S^*)\big)\right|$, and consequently, on the maximum number of failures tolerable by the repair groups in $\C(\bals,S^*)$ for every $\bals\in \fql$, are stated in Theorem~\ref{theo:MHS-bounds}. 
Note that as $q\to \infty$, the two bounds match asymptotically.
In general, it is known that the size of a MHS of a collection of sets can only be approximated to a ratio of $\log$ of the size of the underlying set, i.e. $\log(q^\ell)=\ell\log q$ (see, e.g. \cite{LundYannakakis1994}, \cite{Feige1998}).
Determining the exact size of a MHS of $\C(S^*)$ for an arbitrary subspace $S$ is an open problem.

\begin{theorem}
\label{theo:MHS-bounds}
Let $S$ be a $\delta$-dimensional $\fq$-subspace of $\fql$. Then \vspace{-5pt}
\[
\left\lceil\frac{q^\ell - 1}{q^\delta -1}\right\rceil 
\leq
\left|\mhs\big(\C(S^*)\big)\right| 
\leq
\frac{q^{\ell-\delta +1}-1}{q-1}.\vspace{-5pt}
\]
As a consequence, for every $\bals \in \fql$,\vspace{-5pt}
\[
\left\lceil\frac{q^\ell - 1}{q^\delta -1}\right\rceil-1 
\leq
\left|\F\big(\C(\bals, S^*)\big)\right| 
\leq
\frac{q^{\ell-\delta +1} -1 }{q-1}-1.\vspace{-5pt}
\]
\end{theorem}
\begin{proof} 
The theorem follows directly from Corollary~\ref{cr:UB}, Lemma~\ref{lem:max-failures}, and Lemma~\ref{lem:LB}.
\end{proof}


We can determine $\left|\mhs\big(\C(S^*)\big)\right|$ explicitly for a few special cases when the two bounds collapse.

\begin{corollary}
\label{cor:mhs-subfields}
Let $S$ be a multiplicative coset of the subfield $\ff_{q^\delta}$ in $\fql$. Then, $\left|\mhs\big(\C(S^*)\big)\right| = \frac{q^\ell - 1}{q^\delta -1}$. 
\end{corollary}

\begin{proof}
We consider an equivalence relation $\sim$ over the multiplicative group $\fqls$ as follows:
for $a, b \in \fqls$, $a \sim b$ if and only if $b^{-1}a\in \bbF^*_{q^\delta}$. This relation classifies $\fqls$ into $\frac{q^\ell -1}{q^\delta - 1}$ disjoint equivalence classes, which are the multiplicative cosets of $\ff_{q^\delta}$ in $\fql$. 
Since $\C(S^*) = \C(b\bbF^*_{q^\delta}) = \C(\bbF^*_{q^\delta})$, for all $b \in \fqls$, and
the equivalence classes are disjoint, $\left|\mhs\big(\C(S^*)\big)\right| = |\C(\bbF^*_{q^\delta})| = \frac{q^\ell -1}{q^\delta - 1}$.
\end{proof}

\begin{corollary}
\label{cor:ell-minus-one-dim}
Assume that $(\ell-\delta) \mid {\sf{gcd}}(\ell,\delta)$ and that $S$ is an $\ff_{q^{\ell-\delta}}$-subspace of $\fql$ of dimension $\delta/(\ell-\delta)$ . Then $\left|\mhs \big(\C(S^*)\big)\right| = q^{\ell-\delta} + 1$.
\end{corollary}
\begin{proof}
Let $\overline{q} \define q^{\ell-\delta}$, $\overline{\ell} \define \ell/(\ell-\delta)$ and $\overline{\delta} \define \delta/ (\ell-\delta)$. Replacing $q$, $\ell$, $\delta$ in Theorem~\ref{theo:MHS-bounds} by $\overline{q}$, $\overline{\ell}$, and $\overline{\delta}$, both left and right hand sides are equal to $\overline{q}+1$ or $q^{\ell-\delta}+1$, noting that $\overline{\ell}-\overline{\delta}=1$.
\end{proof}

\begin{example}
\label{ex:mhs-coset-of-field}
Consider the code RS$(16,2)$ over $\bbF_{2^4}$. The set $S = \{0, z^2, z^7, z^{12}\}$, where $z$ is a primitive element of $\bbF_{2^4}$, is a coset of the subfield $\bbF_{2^2}$, and have $\left|\mhs\big(\C(S^*)\big)\right| = |\C(\bbF_{2^2})| = 5$. We can generate five repair schemes for $f(\bals)$ from $S^*$, which tolerate \textit{four} node failures: when four nodes fail, at least one of the five sets of helper nodes survives. For example, when $\bals = z^5$, the five repair schemes correspond to five evaluation point sets in $\C(z^5,S^*) =$ $\big\{\{z, z^{13}, z^{14}\},$ $\{z^{11}, z^4, z^7\},$ $\{z^8, z^6, z^{12}\},$ $\{z^{10}, 0, 1\},$ $\{z^2, z^9, z^{3}\}\big\}$.
If $S = \{0, z^4, z^5, z^8\}$ is chosen, however, then $\left|\mhs\big(\C(S^*)\big)\right| = 6$, and fifteen repair schemes corresponding to fifteen evaluation point sets of $\C(z^5,S^*)$ can be generated. Together, they can tolerate \textit{five} node failures.
\end{example}


\vspace{-3pt}
\subsection{Compact Repair Groups from Multiple Seeds}
\vspace{-3pt}

Although the single-seed compact repair groups discussed in the previous section can \textit{asymptotically} achieve the upper bound on the number of maximum tolerable failures, when $q$ is small, the gap can still be significant.
In this section we discuss the multi-seed approach and determine a set of seeds that attain the upper bound stated in Theorem~\ref{theo:MHS-bounds}. We still assume that the seeds are subspaces of $\fql$. The case of general (non-subspace) seeds is left for future research.

Suppose we build compact repair groups $\C\big(\{S^*_t\}_{t \in [s]}\big)$ from $s$ subspace-seeds $\{S_t\}_{t \in [s]}$. 
If $\dim_{\fq}(S_t) = \delta$ for all $t \in [s]$ then $\C\big(\{S^*_t\}_{t \in [s]}\big) \subseteq \Sds$ and the upper bound in Theorem~\ref{theo:MHS-bounds} still holds.
That is, $\left|\mhs\big(\C\big(\{S^*_t\}_{t \in [s]}\big)\big)\right| \leq \left|\mhs\big(\Sds\big)\right|$.
Therefore, let $s$ be such that $\cup_{t \in [s]}\C\big(\{S^*_t\}_{t \in [s]}\big)\big) = \Sds$ then the repair groups based on these $s$ seeds will attain the upper bound $\left|\mhs\big(\Sds\big)\right|$. Note that this is a sufficient but not necessary condition for $s$.
Finding a smallest $s$ remains an open problem.

\vspace{-5pt}
\begin{lemma}
\label{lem:orbits}
Let $s \define |\Sds/\fqls|$ be the number of orbits considering the action of the group $\fqls$ on $\Sds$ with the standard field multiplication. Let $\{S^*_t\}_{t \in [s]}\subseteq \Sds$ correspond to the set of $s$ disjoint orbits. Then $\left|\mhs\big(\C\big(\{S^*_t\}_{t \in [s]}\big)\big)\right| = \left|\mhs\big(\Sds\big)\right|$.
\end{lemma}
\begin{proof}  
It is clear that the \textit{orbit} of each $S^*\in \Sds$ is $\{bS^*\colon b \in \fqls\}\equiv\C(S^*)$. Therefore, $\Sds$ can be written as the disjoint union of $s$ cosets $\C\big(\{S^*_t\}_{t \in [s]}\big)$, which explains the lemma.
\end{proof} 
\vspace{-5pt}

Armed with Lemma~\ref{lem:orbits}, our next task is to calculate the number of orbits $|\Sds/\fqls|$ explicitly. We acomplish this with Burnside's Lemma and the M\"obius inversion formula.
We recall that for $S^* \in \Sds$, the {\textit{stabilizer} of $S^*$ in $\fqls$ is the set
$\stabS \triangleq \{b \in \fqls: bS^* = S^*\}$.

\vspace{-5pt}
\begin{lemma}[Burnside's Lemma]\cite{Smith2009}
\label{lem:Burnside}
$
|\Sds/\fqls| = \frac{1}{|\fqls|}
\sum_{S^* \in \Sds}|\emph{Stab}_{\fqls}(S^*)|$.
\end{lemma}
\vspace{-5pt}

We first need to determine $\stabS$ for $S^*\in \Sds$.

\vspace{-5pt}
\begin{definition}
\label{def:base}
Let $S$ be an $\fq$-subspace of $\fql$ and $m$ be the largest positive integer so that $S$ is an $\fqm$-subspace of $\fql$. We say that $S$ has \textit{base} $\fqm$ and write $\base(S) = \fqm$.
\end{definition}
\vspace{-5pt}


\vspace{-5pt}
\begin{lemma}
\label{lem:stab-base}
Let $S$ be a subspace of $\fqls$ with base $\fqm$. Then $\stabS = \fqm^*$. Hence, $|\stabS| = q^m-1$.
\end{lemma}
\begin{proof}
Assume that $S$ is a subspace of $\fqls$ with base $\fqm$. Since $S$ is an $\fqm$-subspace, $\alpha S = S$ for all $\alpha \in \fqm^*$. Therefore, $\fqm^*\subseteq \stabS$.
It suffices to show that the opposite is also true, i.e. if $\alpha S = S$ for $\alpha \in \fql$ then $\alpha\in \fqm^*$. Indeed,
assume that $\alpha \notin \fqm^*$ for the sake of contradiction. Set $\bbF = \fqm(\alpha)\supsetneq \fqm$, which is the extension field of $\fqm$ with respect to $\alpha$.
Then $\bbF = \left\{\sum_{i=0}^{\Delta-1}c_i\alpha^i: c_i \in \fqm\right\}$, where $\Delta$ is the degree of a minimal polynomial of $\alpha$ over $\fqm$.
Since $\alpha^i u \in S$ for every $i\geq 0$ and $u \in S$, we deduce that
$\big(\sum_{i=0}^{\Delta-1}c_i\alpha^i\big)u = \sum_{i=0}^{\Delta-1}c_i(\alpha^i u) \in S$.
Hence, $S$ is also an $\bbF$-subspace of $\fql$, where $\bbF$ is larger than $\fqm$, which is a contradiction to the assumption that $\fqm$ is the base of $S$. Hence, $\alpha \in \fqm^*$. 
\end{proof}

\vspace{-5pt}
From Lemma~\ref{lem:stab-base}, we know that the stabilizer of $S^* \in \Sds$ with respect to $\fqls$ is $\base(S)$.
To apply Burnside's Lemma, it remains to count the number of $S^* \in \Sds$ that have base $\fqm$ for each $m \mid \gcd(\ell,\delta)$. Note that the $q$-ary Gaussian coefficient\vspace{-5pt}
\[
 \left[
\begin{matrix}
\ell  \\
\delta \\
\end{matrix}
\right]_q
\triangleq
\frac{(q^\ell - 1)(q^\ell - q)(q^\ell - q^2)\dots (q^\ell - q^{\delta -1})}
{(q^\delta - 1)(q^\delta - q)(q^\delta - q^2)\dots (q^\delta - q^{\delta -1})}\vspace{-5pt}
\]
only counts the number of $\fq$-subspaces of $\fql$ of dimension $\delta$, among which are subspaces of different bases. For instance, with $\ell = 48$, $\delta = 24$, $\gcd(\ell, \delta) = 24$ and its divisors are $2, 3, 4, 6, 8, 12, 24$. 
A $24$-dimensional $\fq$-subspace can have base
$\bbF_{q^2}$, $\bbF_{q^3}$, $\bbF_{q^4}$, $\bbF_{q^6}$, $\bbF_{q^8}$, $\bbF_{q^{12}}$, or $\bbF_{q^{24}}$.
Although the Gaussian coefficient doesn't give us directly the number of $\delta$-dimensional $\fq$-subspaces of $\fql$ that have base $\fqm$, it can still be used to derive this number via the M\"obius inversion formula.




The \textit{M\"obius function} $\mu(v)$ (see~\cite[Chapter XVI]{HardyWright1975}) is defined over the positive integers as follows: $\mu(v)\triangleq 0$ if $v$ is not square-free, and $(-1)^r$ if $v$ is the product of $r$ distinct primes. 

\begin{lemma}[M\"obius inversion formula]\cite[Chap.~XVI]{HardyWright1975}
\label{lem:22}
Let $f$ and $g$ be functions over $\bbZ^+$ and 
$f(n) = \sum_{v|n}g(v).$
Then, $g(n) = \sum_{v|n}\mu(v)f\big(n/v\big)$,
where $\mu(v)$ is the M\"obius function.
\end{lemma}


\begin{theorem}
\label{theo:orbits}
Let $\nqld(m)$ be the number of $\delta$-dimensional $\fq$-subspaces of $\fql$ with base $\fqm$. 
Let $(a,b)\define {\sf{gcd}}(a,b)$.
Then \vspace{-5pt} 
\begin{equation}
\label{eq:1}
\nqld(m)
= \sum\nolimits_{v|\frac{(\ell, \delta)}{m}}
\mu(v)
\left[
\begin{smallmatrix}
\ell/mv \\
\delta/mv \\
\end{smallmatrix}
\right]_{q^{mv}}.\vspace{-5pt}
\end{equation}
Using Burnside's Lemma, we have \vspace{-5pt}
\begin{equation}
\label{eq:2}
|\Sds/\fqls| \hspace{-2pt}
=\hspace{-2pt} \dfrac{1}{q^\ell\hspace{-2pt}-\hspace{-2pt}1}\hspace{-2pt}
\sum_{m|{(\ell, \delta)}}\hspace{-2pt}(q^m-1)\bigg(\hspace{-2pt}\sum_{v|\frac{(\ell, \delta)}{m}}\hspace{-2pt}
\mu(v)\hspace{-2pt}
\left[
\begin{smallmatrix}
\ell/mv \\
\delta/mv \\
\end{smallmatrix}
\right]_{q^{mv}}\hspace{-2pt}\bigg)\hspace{-1pt}. \vspace{-5pt}
\end{equation}
\end{theorem}
\begin{proof}
To prove \eqref{eq:1}, we note that the Gaussian coefficient 
$
\left[
\begin{smallmatrix}
\ell/m \\
\delta/m \\
\end{smallmatrix}
\right]_{q^m}
$
counts the number of $\delta/m$-dimensional $\fqm$-subspaces of $\fql$, which include all subspaces with bases $\bbF_{q^p}$ where $p \geq m$, $m\mid p$, and $p \mid (\ell,\delta)$.
Hence,\vspace{-5pt}
\begin{equation}
\label{eq:3}
\left[
\begin{smallmatrix}
\ell/m \\
\delta/m \\
\end{smallmatrix}
\right]_{q^m}
=\sum\nolimits_{p \geq m,~m \mid p,~ p|(\ell, \delta)}
\nqld(p).\vspace{-5pt}
\end{equation}

Fixing $q$, $\ell$, and $\delta$, for each $m$ let 
$n = (\ell,\delta)/m$
and
$v = (\ell,\delta)/p$.
Since $m \mid p$, we also have $v \mid n$.
It is obvious that
$m = (\ell,\delta)/n$
and
$p = (\ell,\delta)/v$.
Define\vspace{-5pt}
\[
f(n) 
\triangleq
\left[
\begin{smallmatrix}
\ell/m \\
\delta/m \\
\end{smallmatrix}
\right]_{q^m}
=
\left[
\begin{smallmatrix}
\ell n/(\ell,\delta) \\
\delta n/(\ell, \delta) \\
\end{smallmatrix}
\right]_{q^{\frac{(\ell,\delta)}{n}}},\vspace{-5pt}
\]
\[
g(v)
\triangleq
\nqld(p) 
=
\nqld\Big(\frac{(\ell,\delta)}{v}\Big).\vspace{-5pt}
\]
From \eqref{eq:3}, we have 
$f(n) = \sum_{v|n} g(v)$.
Applying the M\"obius inversion formula we obtain
$
g(n) = \sum_{v|n}\mu(v)f(n/v),
$
or\vspace{-5pt}
\[
\begin{split}
\nqld(m) = 
\nqld
\left(
\frac{(\ell, \delta)}{n}
\right)
&=
\sum_{v|n}\mu(v)
\left[
\begin{smallmatrix}
\ell \frac{n}{v}/(\ell,\delta) \\ \vspace{-5pt}
\delta \frac{n}{v}/(\ell, \delta) \\
\end{smallmatrix}
\right]_{q^{\frac{(\ell,\delta)}{\frac{n}{v}}}}\\ 
&= \sum_{v|\frac{(\ell, \delta)}{m}}
\mu(v)
\left[
\begin{smallmatrix}
\ell/mv \\
\delta/mv \\
\end{smallmatrix}
\right]_{q^{mv}},
\end{split}\vspace{-15pt}
\]
which establishes \eqref{eq:1}.

From Burnside's Lemma and Lemma~\ref{lem:stab-base}, we have \vspace{-5pt}
\[
\begin{split}
|\Sds/\fqls| &= \frac{1}{|\fqls|} \sum_{S^* \in \Sds}|\stabS|\\ \vspace{-5pt}
&= \frac{1}{q^\ell-1}\sum_{m \mid (\ell,\delta)}\hspace{5pt} \sum_{S^* \in \Sds,~\base(S^*) = \fqm^*} (q^m-1)\\ \vspace{-10pt}
&= \frac{1}{q^\ell-1}\sum_{m \mid (\ell,\delta)} (q^m-1)\nqld(m),
\end{split}\vspace{-10pt}
\]
which implies \eqref{eq:2}.
\end{proof}
\vspace{-5pt}

By using Theorem \ref{theo:orbits}, we achieve an upper bound on the minimum number of seeds that generate a collection of repair groups attaining the upper bound in Theorem~\ref{theo:MHS-bounds}. Identifying the exact value remains an open problem.

\vspace{-5pt}
\begin{corollary}
\label{cor:UB-subspaces}
Let $s = |\Sds/\fqls|$ as given in Theorem~\ref{theo:orbits} be the number of orbits with respect to the action of $\fqls$ on $\Sds$, and $\{S^*_t\}_{t \in [s]}\subseteq \Sds$ correspond to the set of $s$ disjoint orbits. Then $\left|\mhs\big(\C\big(\{S^*_t\}_{t \in [s]}\big)\big)\right| =
\frac{q^{\ell - \delta +1}-1}{q-1}$.
Using these $s$ seeds, the resulting compact repair groups can tolerate $\F\big(\C\big(\bals,\{S^*_t\}_{t \in [s]}\big)\big) = \frac{q^{\ell - \delta +1}-1}{q-1}-1$ failures.
\end{corollary}
\textbf{Acknowledgement.} This work is supported by the ARC DECRA Grant DE180100768 and ARC DP200100731. 

\newpage
\bibliographystyle{IEEEtran}
\bibliography{MultipleRepairGroups}

\end{document}